\documentclass[lettersize,journal]{IEEEtran}
\usepackage{amsmath,amsfonts}
\usepackage{array}
\usepackage[caption=false,font=normalsize,labelfont=sf,textfont=sf]{subfig}
\usepackage{textcomp}
\usepackage{stfloats}
\usepackage{url}
\usepackage{float}
\usepackage{verbatim}
\usepackage{multirow}
\usepackage{graphicx}
\usepackage{booktabs}
\usepackage{algpseudocode}
\usepackage{diagbox}
\usepackage{tabularx}
\usepackage{indentfirst}
\usepackage{algorithmicx,algorithm}
\usepackage[justification=centering]{caption}
\def\BibTeX{{\rm B\kern-.05em{\sc i\kern-.025em b}\kern-.08em
    T\kern-.1667em\lower.7ex\hbox{E}\kern-.125emX}}
\usepackage{balance}
\begin{document}
\title{Large-Scale Traffic Signal Control by a Nash Deep Q-network Approach}
\author{Yuli Zhang, Shangbo Wang, Ruiyuan Jiang }

\markboth{}%
{How to Use the IEEEtran \LaTeX \ Templates}

\maketitle

\begin{abstract}
Reinforcement Learning (RL) is currently one of the most commonly used techniques for traffic signal control (TSC), which can adaptively adjusted traffic signal phase and duration according to real-time traffic data. However, a fully centralized RL approach is beset with difficulties in a multi-network scenario because of exponential growth in state-action space with increasing intersections. Multi-agent reinforcement learning (MARL) can overcome the high-dimension problem by employing the global control of each local RL agent, but it also brings new challenges, such as the failure of convergence caused by the non-stationary Markov Decision Process (MDP). In this paper, we introduce an off-policy nash deep Q-Network (OPNDQN) algorithm, which mitigates the weakness of both fully centralized and MARL approaches. The OPNDQN algorithm solves the problem that traditional algorithms cannot be used in large state-action space traffic models by utilizing a fictitious game approach at each iteration to find the nash equilibrium among neighboring intersections, from which no intersection has incentive to unilaterally deviate. One of main advantages of OPNDQN is to mitigate the non-stationarity of multi-agent Markov process because it considers the mutual influence among neighboring intersections by sharing their actions. On the other hand, for training a large traffic network, the convergence rate of OPNDQN is higher than that of existing MARL approaches because it does not incorporate all state information of each agent.  We conduct an extensive experiments by using Simulation of Urban MObility simulator (SUMO), and show the dominant superiority of OPNDQN over several existing MARL approaches in terms of average queue length, episode training reward and average waiting time. \\
\end{abstract}


\section{Introduction}
\IEEEPARstart{W}{ith} the growth of vehicle ownership, the current traffic demand is rising rapidly. Traffic jams and vehicular accidents are becoming more seriously with the increasing travel demand. One of the main reasons is the unreasonable traffic signal settings. It has been reported in \cite{Liang2018} that the current traffic signal network cannot adaptively respond to the real time traffic state. Existing traffic light control either deploys fixed-time mode without considering real-time traffic or considering the traffic to a very limited degree \cite{casas2017deep}, , which cause the vehicle accumulation at intersections and degrade traffic efficiency \cite{Li2020}. Therefore, developing an effective traffic signal control strategy is of great significance in urban network. 

In recent years, reinforcement learning (RL) technique had many significant achievements in traffic signal control (TSC) because RL can better solve the decision optimization problem of sequential actions. In the literature, TSC using RL methods could be categorized into model-based and model-free approaches. In a model-based approach, agent trying to understand environment and creating a model based on interact with environment. A model-free RL approach learns the optimal policy base without thoroughly understanding the dynamics of the traffic system \cite{jin2019multi}. A model-free method is straightforward and convenient for implementation, therefore, model-free RL methods are widely used in TSC. There are two variants for model-free RL: Value-based and Policy-based method. Q-learning is a value-based RL method which able to compare the expected utility of the available actions without requiring a model of the environment. However, the Q-table will be huge if state space is vast, and searching and storing it involves a lot of time and memory, so Q-learning approach performance depends on the quality of traffic feature design \cite{Chu2015}, \cite{Tan2018}, \cite{abdoos2011traffic}. Deep Q-Network (DQN), first proposed in 2015 by Mnih et al., successfully combines RL with deep learning by using a multi-layer convolutional neural network to approximate the Q-function \cite{Mnih2015a}. It can be observed from \cite{Chu2016}, \cite{Aragon-Gomez2020} that DQN can achieve good performance for small state-action space in single-agent control. In 2019, Liang and Du proposed a Double-Dueling-Deep Q-Network method, including dueling network, double network and deep network \cite{Liang2019} to control the traffic light cycle in a single intersection.

Even though RL technique has made good achievements in TSC for single intersection scenario, it is infeasible for large-scale urban road network with multiple intersections because of exponential growth of state-action space with increase of number of intersections. To overcome the extremely high dimension of joint action space issue, some researchers proposed multi-agent reinforcement learning (MARL) approach, which distributes global control to each local agent \cite{prabuchandran2014multi}, \cite{wang2020large}. In 2020, Chu et al., \cite{Chu2020} proposed an Advantage Actor-Critic (A2C) algorithm in an adaptive TSC environment, which can improve the observability of each agent by considering the control policies of other agents. Chen et al., improved A2C by using difference reward method \cite{Chen2021} in 2021. However, the disadvantage of the distributed A2C approach is that when the state dimension or the number of surrounding intersections increases, intersections will have a severe mutual influence, resulting in a non-stationary environment.

Since Nash presented the concept of nash equilibrium in game theory in \cite{nash1995}, there are some research about nash Q-learning. In 2004, Hu and Wellman designed an on-policy nash Q-learning algorithm which extended Q-learning to non-cooperative multi-agent environments \cite{Hu2004}. However, this on-policy nash Q-learning algorithm has an apparent defect in that it needs to know or be able to calculate the control policies of each agent. We conclude that the on-policy approach is impractical in complex traffic networks, because it is cumbersome to obtain the accurate control policies. To tackle the problem, this paper proposes an off-policy nash deep Q-Network (OPNDQN) algorithm that find the optimal strategy by searching the nash equilibrium among neighboring intersections iteratively without need of knowing or calculating the control policy, i.e. the conditional probability of selecting an action based on current state. The main contributions are as follows:
\begin{itemize}
	\item[$\bullet$] We propose the OPNDQN algorithm that can achieve the nash equilibrium of the expected cumulative reward for non-cooperative agents without knowing or calculating the control policies, and apply it in a large-scale traffic signal control scenario;  
	\item[$\bullet$] We evaluate the performance of the proposed approach, by conducting extensive experiments using SUMO. (Simulation of Urban MObility) \cite{krajzewicz2012recent}. We compare the OPNDQN algorithm to Fully Decentralized DQN approach, Multi-agent Q-Learning (MAQL), Cooperative Multi-agent Deep Reinforcement Learning approach (Co-MARL) \cite{HADDAD2022105019} and Multi-agent Advantage Actor-Critic (MA2C). The results show that the proposed OPNDQN approach outperforms these methods in terms of multiple traffic metrics that are: average queue length, episode training reward, and average waiting time. 
\end{itemize} 
\section{METHODOLOGY}
To let reader better understand the OPNDQN approach, we firstly briefly review the Markov Decision Process.
\subsection{Markov Decision Process}
Markov Decision Process (MDP) is the basic framework for illustrating RL problem, which is described by a sequential decision process that state space, action space, reward, state transition function, and discount factor. RL is a sequential decision process that interacts with the environment to maximize a target reward function through long-term trial and error. A MDP can be defined with a quin-tuple $\langle \textit{S}, \textit{A}, \textit{R}, \textit{T},\gamma\rangle$:
\begin{itemize}
	\item[$\bullet$]State space $\textit{S}$: The set of all possible states. State is the description and generalization of the environment. $ \textit{s}_{\textit{t}} $ $\left(\textit{s}_{\textit{t}} \in \textit{S}\right)$ is a state observed by an agent at $ \textit{t} $-th time instant. 
	\item[$\bullet$]Action space $\textit{A}$: The set of all possible actions. $ \textit{a}_{\textit{t}} $ $\left(\textit{a}_{\textit{t}} \in \textit{A}\right)$ is an action of an agent at $ \textit{t} $-th time instant.  
	\item[$\bullet$]Reward $\textit{R}$: The reward space. After the agent performs an action, the environment returns a value (reward) to the agent. $ \textit{r}_{\textit{t}} $ $\left(\textit{r}_{\textit{t}} \in \textit{R}\right)$ is a reward of an agent at $ \textit{t} $-th time instant.   
	\item[$\bullet$]State transition function $\textit{T}$: A function used by the environment to generate new states. It represents the probability of transitioning from one state to another. The state transition function $ \textit{T} $ can be expressed as: $p\left(\textit{s}_{\textit{t}+1} \mid \textit{s}_{\textit{t}},\textit{a}_{\textit{t}}\right)=\mathbb{P}\left(\textit{S}^{\prime}=\textit{s}_{\textit{t}+1} \mid \textit{S}=\textit{s}_{\textit{t}}, A=\textit{a}_{\textit{t}}\right)$, where $ \textit{p} $ is the probability, $ \textit{a}_{\textit{t}} $ is the action of an agent at $ \textit{t} $-th time instant, $ \textit{s}_{\textit{t}} $ is the state of an agent, and $ \textit{s}_{\textit{t}+1} $ represents the state that all agents transfer to at the next moment. 
	\item[$\bullet$]Discount factor $\gamma$: Discounts on future reward. $(\gamma \in[0,1])$ 
\end{itemize}

\subsection{OPNDQN}
 To facilitate the reader's understanding of the OPNDQN method, we will start with a single-agent system. Let us define $ \textit{s}_{\textit{t}} $, $ \textit{a}_{\textit{t}} $, $ \textit{r}_{\textit{t}} $ as the state matrix, action vector and reward vector at $ \textit{t} $-th time instant, respectively, in a MDP, we define all rewards from start to end as: $\textit{r}_{1}, \cdots, \textit{r}_{\textit{t}}, \cdots, \textit{r}_{t_{\textit{End}}} $. Combine discount rates, future discounted return at time $ \textit{t}$ as $ \textit{U}_{\textit{t}} $, it can be represented as:\\
\begin{equation}
\textit{U}_{\textit{t}}=\textit{r}_{\textit{t}}+\gamma \cdot \textit{r}_{\textit{t}+1}+\gamma^{2} \cdot \textit{r}_{\textit{t}+2}+\cdots+\gamma^{\textit{t}_{\textit{End}}-\textit{t}} \cdot \textit{r}_{\textit{t}_{\textit{End}}}
\label{1}
\end{equation}
The randomness of $ \textit{U}_{\textit{t}} $ comes from all sequences of actions and observations after time $ \textit{t} $. To eliminate the effect of randomness, we use the expectation of Equation (1) and define the action-value function $ \textit{Q}_{\pi}\left(\textit{s}_{\textit{t}}, \textit{a}_{\textit{t}}\right) $ as\\
\begin{equation}\label{2}
\textit{Q}_{\pi}\left(\textit{s}_{\textit{t}}, \textit{a}_{\textit{t}}\right)=\mathbb{E}\left[\textit{U}_{\textit{t}} \mid \textit{S}_{\textit{t}}=\textit{s}_{\textit{t}}, \textit{A}_{\textit{t}}=\textit{a}_{\textit{t}},\pi\right]
\end{equation}
where $ \pi $ is a policy mapping sequences to actions. Traditional DQN method uses the maximization method to determine the optimal policy $ \pi^{*} $ and derive the  optimal action-value function:\\
\begin{equation}\label{3}
\textit{Q}^{*}(\textit{s}_{\textit{t}}, \textit{a}_{\textit{t}})=\max _{\pi} \mathbb{E}\left[\textit{U}_{\textit{t}} \mid \textit{S}_{\textit{t}}=\textit{s}, \textit{A}_{\textit{t}}=\textit{a}, \pi^{*}\right]
\end{equation}
The optimal action-value function obeys an important identity known as the \textit{Bellman equation} \cite{Mnih2015a}. Assuming that the agent knows the optimal Q-value for the subsequent state, then the optimal policy is to select action $ \textit{a}_{t+1} $ maximising the expected value: $\textit{r}_{\textit{t}}+\gamma \textit{Q}^{*}\left(\textit{s}_{\textit{t}+1}, \textit{a}_{\textit{t}+1}\right)$. In practice, it is common to use a neural network to approximate the Q-value:\\
\begin{equation}\label{4}
\textit{Q}(\textit{s}_{\textit{t}}, \textit{a}_{\textit{t}} ; \theta) \approx \textit{Q}^{*}(\textit{s}_{\textit{t}}, \textit{a}_{\textit{t}})
\end{equation}
where $ \theta $ is the parameter of the neural network. We define a neural network function approximator as a Q-network. 

However, in MARL, multi-agent system interacts with the same environment, which make the neural network unable to converge because of non-stationarity. To fill the gap, we try to investigate how to mitigate the non-stationarity of the MDP by considering the influence caused by neighboring agents. Some of literature tends to let each agent collect the information of all other agents to increase the observability in the process of model training to increase the stationarity, however, it will negatively affect the convergence rate because of extremely large dimension of state-action space \cite{HADDAD2022105019}. In this paper, we propose an OPNDQN approach, which finds the nash equilirium strategies among neighboring agents at each iteration, by only sharing their actions. The advantage of this approach is not simply to increase the Q-value of each agent, but to find a balance between the agent and its neighbors so that each agent increases its own rewards without lowering the rewards of neighbors. We consider a non-cooperative multi-agent system with $ \textit{N} $ agents, each of which has $ \textit{n} $ neighbors. For all $\textit{s}^{1} \in \textit{S}^{1}, \ldots, \textit{s}^{\textit{N}} \in \textit{S}^{\textit{N}}, \textit{a}^{1} \in \textit{A}^{1}, \ldots, \textit{a}^{\textit{N}} \in \textit{A}^{\textit{N}}$, we use a fictitious game method to find the nash Q-value for complex model-free learning of multi-agent. We define that the action selection by the agent $ \textit{i} $ ($ \textit{i} \in \textit{N} $) depends not only on its own state, action and reward values but also on the action of their neighbours $ \hat{\textit{i}} $ ($ \hat{\textit{i}} \in \textit{n} $):
\begin{equation}\label{5}
	\textit{a}_{\textit{t}}^{\textit{i}}=\textit{Nash} \textit{Q}_{\textit{t}}\left(\textit{s}_{\textit{t}}^{\textit{i}}, \textit{a}_{\textit{t}}^{\textit{i}},\textit{a}_{\textit{t}}^{\hat{\textit{i}}} ; \theta_{\textit{i}}\right)
\end{equation}
The fictitious game process can be described as follows: initialize the actions $ \textit{a}_{\textit{t}}^{\textit{i}} $ and $ \textit{a}_{\textit{t}}^{\hat{\textit{i}}} $ at each episode, then update the group actions to increase each agent's Q-value to the Nash Equilibrium, where no agent can further increase its Q-value by unilaterally change its action (Eq. 6).
\begin{equation}\label{6}
	\textit{Nash} \textit{Q}^\textit{i}\left(s^\textit{i}, a^\textit{i}, a^{\hat{\textit{i}}} ; \theta_\textit{i}\right)_{\text {Step } \textit{j}} \le \textit{Nash} \textit{Q}^\textit{i}\left(s^\textit{i}, a^\textit{i}, a^{\hat{\textit{i}}} ; \theta_\textit{i}\right)_{\text {Step } \textit{j}-1}
\end{equation}
where $ \textit{j} $ is fictitious game loop counter. The action (real action to do) with the largest Q-value among all the previously selected actions is the nash equilibrium strategy. It is worth noting that in the above fictitious game process, the process of a searching for nash equilibrium by changing actions $ a_{t}^{\hat{i}} $ is fictitious, while the agent $ \textit{i} $ and its neighbor $ \hat{\textit{i}} $ do not make real action. The agents does not make real actions until the final nash strategy $ \textit{a}^{ \textit{i}}_{\text{Step }\textit{j}} $ is found. The flow of the fictitious game can be seen in Fig. 1. Fictitious game in pseudocode of OPNDQN is as follows:
\begin{algorithm}[H]
	\caption{Fictitious Game} 
	\hspace*{0.02 in} {\bf Input:} 
	random actions $ \textit{a}^{\textit{i}} $\\
	\hspace*{0.02 in} {\bf Output:}  
	nash joint actions $ \textit{a}^{\textit{i}}_{\text{Step } \textit{j}}$ or random actions $ \textit{a}^{\textit{i}} $\\
	\hspace*{0.02in} {\bf Notation:} The parameters in the neural network $\theta$,
	agents \hspace*{0.075 in}number $\textit{n}$,
	nash parameter $\epsilon$
	\begin{algorithmic}[1]
		\State Initialize $ \textit{a}^{\textit{i}} $ to the random action.
		\If{probability $\epsilon$ select random action}
		\State Output $ \textit{a}^{\textit{i}} $ 
		\Else 
		\While{Step $ \textit{j}=1 $}
		\State $ \textit{a}^{ \textit{i}}_{\text{Step }\textit{j}}=\operatorname{argmax} \textit{Q}^\textit{i} \left(\textit{s}^\textit{i },\textit{\textit{a}}^{\textit{i}}_{\text{Step } \textit{j}-1},\textit{a}^{\hat{\textit{i}}}_{\text{Step } \textit{j}-1} ; \theta_{\textit{i}}\right) $
		\If {eq (\ref{6})}
		\State break
		\EndIf
		\EndWhile
		\State Output $ \textit{a}^{ \textit{i}}_{ \text{Step }  \textit{j}-1 } $ 
		\EndIf

	\end{algorithmic}
\end{algorithm}  
We define the target $ \textit{y}_{\textit{k}}^{\textit{i}} $ and train a  Q-network by minimising a sequence of loss functions $ \textit{L}_{\textit{k}}^{\textit{i}}\left(\theta_{\textit{k}}\right) $ that changes at each iteration $ \textit{k} $:\\
\begin{equation}\label{7}
\textit{y}_{\textit{k}}^{\textit{i}}=\textit{r}_{\textit{t}}^{\textit{i}}+\gamma \cdot \textit{Nash}\textit{Q}^{\textit{i}}\left(\textit{s}_{\textit{t}+1}^{\textit{i}}, \textit{a}_{\textit{t}+1}^{\textit{i},\hat{\textit{i}}} ; \theta_{\textit{k}}\right)
\end{equation}
\begin{equation}\label{8}
\textit{L}_{\textit{k}}^{\textit{i}}\left(\theta_{\textit{k}}\right)=\mathbb{E}\left[\left(\textit{y}_{\textit{k}}^{\textit{i}}-\textit{Q}_{\textit{t}}^{\textit{i}}\left(\textit{s}_{\textit{t}}^{\textit{i}}, \textit{a}_{\textit{t}}^{\textit{i},\hat{\textit{i}}} ; \theta_{\textit{k}}\right)\right)^{2}\right]
\end{equation}
Compute the gradient of loss $ \textit{\textit{L}} $ with respect to $ \theta $:\\
\begin{equation}
\begin{split}
\label{9}
\nabla_{\theta_{\textit{k}}} \textit{L}_{\textit{k}}(\theta_{\textit{k}})=\mathbb{E}[\textit{r}_{\textit{t}}^{\textit{i}}+\gamma \cdot \textit{Nash}\textit{Q}^{\textit{i}}(\textit{s}_{\textit{t}+1}^{\textit{i}}, \textit{a}_{\textit{t}+1}^{\textit{i},\hat{\textit{i}}} ; \theta_{\textit{k}})\\-\textit{Q}_{\textit{t}}^{\textit{i}}(\textit{s}_{\textit{t}}^{\textit{i}}, \textit{a}_{\textit{t}}^{\textit{i},\hat{\textit{i}}} ; \theta_{\textit{k}})) \nabla_{\theta_{\textit{k}}} \textit{Q}_{\textit{t}}^{\textit{i}}(\textit{s}_{\textit{t}}^{\textit{i}}, \textit{a}_{\textit{t}}^{\textit{i},\hat{\textit{i}}} ; \theta_{\textit{k}})]
\end{split}
\end{equation}
The stochastic gradient descent method is used to optimize the loss function $ \textit{L}_{\textit{k}}^{\textit{i}}\left(\theta_{\textit{k}}\right) $, the parameters $ \theta_{\textit{k}} $ of the neural network converge to a fixed value with the increase of the number of training step $ \textit{k} $.

\subsection{Target Network}
To update the parameters $\theta$ in the neural network, a target value $ \theta^{*} $ is defined to help guide the update process. Using the target network to update the parameters, Equation (\ref{7}) can be written as:
\begin{equation}\label{7}
	\textit{y}_{\textit{k}}^{\textit{i}}=\textit{r}_{\textit{t}}^{\textit{i}}+\gamma \cdot \textit{Nash}\textit{Q}^{\textit{i}}\left(\textit{s}_{\textit{t}+1}^{\textit{i}}, \textit{a}_{\textit{t}+1}^{\textit{i},\hat{\textit{i}}} ; \theta_{\textit{k}}^{*}\right)
\end{equation}
\subsection{Overall Architecture}
\begin{figure*}[t]
	\centering
	\includegraphics[width=1\linewidth]{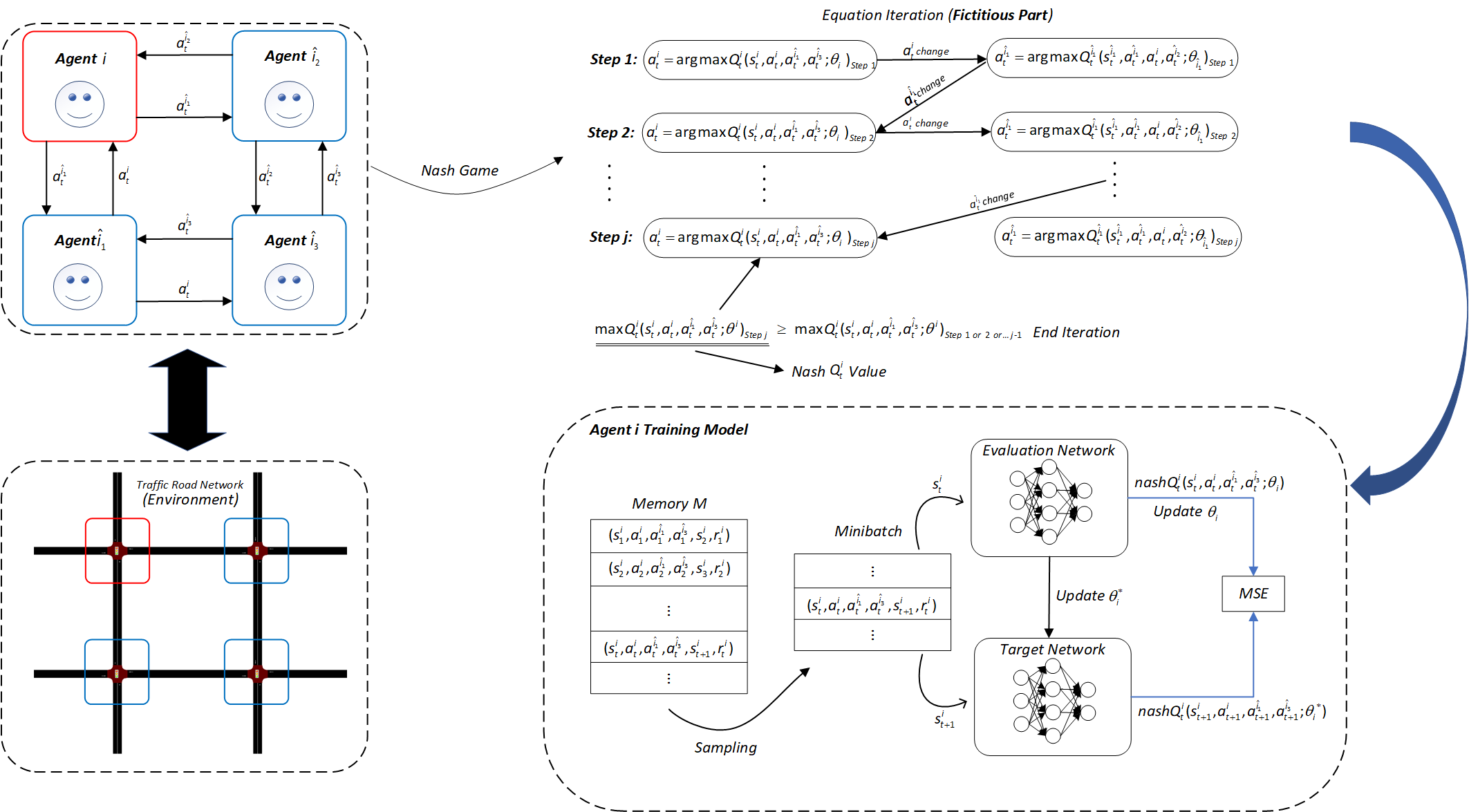}
	\caption{Overall architecture of the OPNDQN approach.}
	\label{chutian22}
\end{figure*}
\begin{algorithm}[H]
\caption{OPNDQN Algorithm} 
\hspace*{0.02 in} {\bf Input:} 
replay memory size $\textit{M}$, minibatch size $\textit{B}$, nash 
\\\hspace*{0.07 in}parameter $\epsilon$, pre-train steps $\textit{tp}$, target network update rate $\alpha$, \\\hspace*{0.07 in}discount  factor $\gamma$.\\
\hspace*{0.02in} {\bf Notation:} The parameters in the primary neural network $\theta$, \hspace*{0.07 in}the parameters in the target neural network $\theta^{*}$, 
the replay \hspace*{0.075 in}memory $\textit{m}$, the agents number $\textit{n}$, training step number $\textit{k}$, \hspace*{0.08in}time $\textit{t}$.
\begin{algorithmic}[1]
	\State Initialize parameters $\theta, \theta^{*}$ with random values.
	\State Initialize $\textit{m}$ to be empty.
	\State Initialize  $\textit{s}^{\textit{i}}$ with the starting scenario at the intersections.
	\State Initialize $ \textit{a}^{\textit{i}} $ to the random action.
	\For{Episode=1}
	\State Initialize state $\textit{s}^{\textit{i}}$ 
	\While{the number of vehicles $ \textit{N}   >  0$, $ k $=1} 
	\State \textbf{Algorithm 1}$ \rightarrow $ $\textit{a}_{\textit{t}}^{\textit{i}}$
	\State Take action $\textit{a}_{\textit{t}}^{\textit{i}}$
	\State observe reward $\textit{r}_{\textit{t}}^{\textit{i}}$ and new state $\textit{s}^{\textit{i}}_{\textit{t}}$,
	\State Add episode to buffer
	\State Assign $\textit{s}_{\textit{t}+1}^{\textit{i}}$ to $\textit{s}_{\textit{t}}^{\textit{i}}: \textit{s}_{\textit{t}}^{\textit{i}} \leftarrow \textit{s}_{\textit{t}+1}^{\textit{i}}$.
	\State $\textit{\textit{k}} \leftarrow \textit{k}+1 .$
	\If{$|\textit{M}|>\textit{B}$ and $\textit{k}>\textit{tp}$}
	\State Select $ \textit{B} $ samples from replay memory $ \textit{m} $
	\State \textbf{Algorithm 1} $ \rightarrow $ $\textit{a}_{\textit{t}+1}^{\textit{i}}$
	\State \textit{nash} $ \textit{Q}^{\textit{i}}_{\textit{t}+1}= \textit{Q}_{\textit{t}+1}^{\textit{i}}\left(\textit{s}_{\textit{t}+1}^{\textit{i}}, \textit{a}_{\textit{t}+1}^{\textit{i}}, \textit{a}_{\textit{t}+1}^{\hat{\textit{i}}} ; \theta^{*}_{\textit{i}}\right) $
	\State Calculate the loss $ \textit{L} $: 
	\State $\begin{aligned} L= \left[\left(\textit{r}+\gamma \textit{nash} \textit{Q}^{\textit{i}}_{\textit{t+1}}\right) -\right.\textit{Q}(\textit{s}_{\textit{t}}^{\textit{i}}, \textit{a}_{\textit{t}}^{\textit{i}}, \textit{a}_{\textit{t}+1}^{\hat{\textit{i}}}; \theta_{\textit{i}})]^{2} . \end{aligned}$
	\State Update $\theta_{\textit{i}}$ with $\nabla L$ using RMS optimizer.
	\State Assign $\theta_{\textit{i}}$ to $\theta^{*}_{\textit{i}}$.
	\If{the number of vehicles $ \textit{N}   =  0$}
	\State break
	\EndIf
	\EndIf
	\EndWhile
	\EndFor
	
\end{algorithmic}
\end{algorithm} 
As illustrated in Fig. 1, we took four intersections as an example to show the training process of an agent $ \textit{i} $ (red box). We can see that the intersection with red box has three neighbors (blue box). The intersections can transmit action information to each other, and the agent can receive the action that the adjacent intersections want to change. After train step number $ \textit{k} $ greater than or equal to the pre-train steps value $ \textit{tp} $ is completed, the parameter of agent $ \textit{i} $ is trained in the training model. We use evaluation network and the target network with the same fully connected network for the DQN used in this paper. Each network contains three hidden layers with sizes of 32, 64, 64, and each layer selects the $ \textit{relu} $ function as the activation function. The output layer outputs the Q-value of the action. Root Mean Square (RMS) with learning rate of $ 0.0001 $ was selected as the gradient optimizer. In training, we utilize a mini-bacth method with a batchsize of 64.

\section{SIMULATION ENVIRONMENT}
In this section, we will describe the applied simulation environment for this research. We will firstly illustrate the road network model, and followed by the definition of states, actions and reward. 
\subsection{Road network}
\begin{figure}[!]
	\centering
	\includegraphics[width=0.8\linewidth]{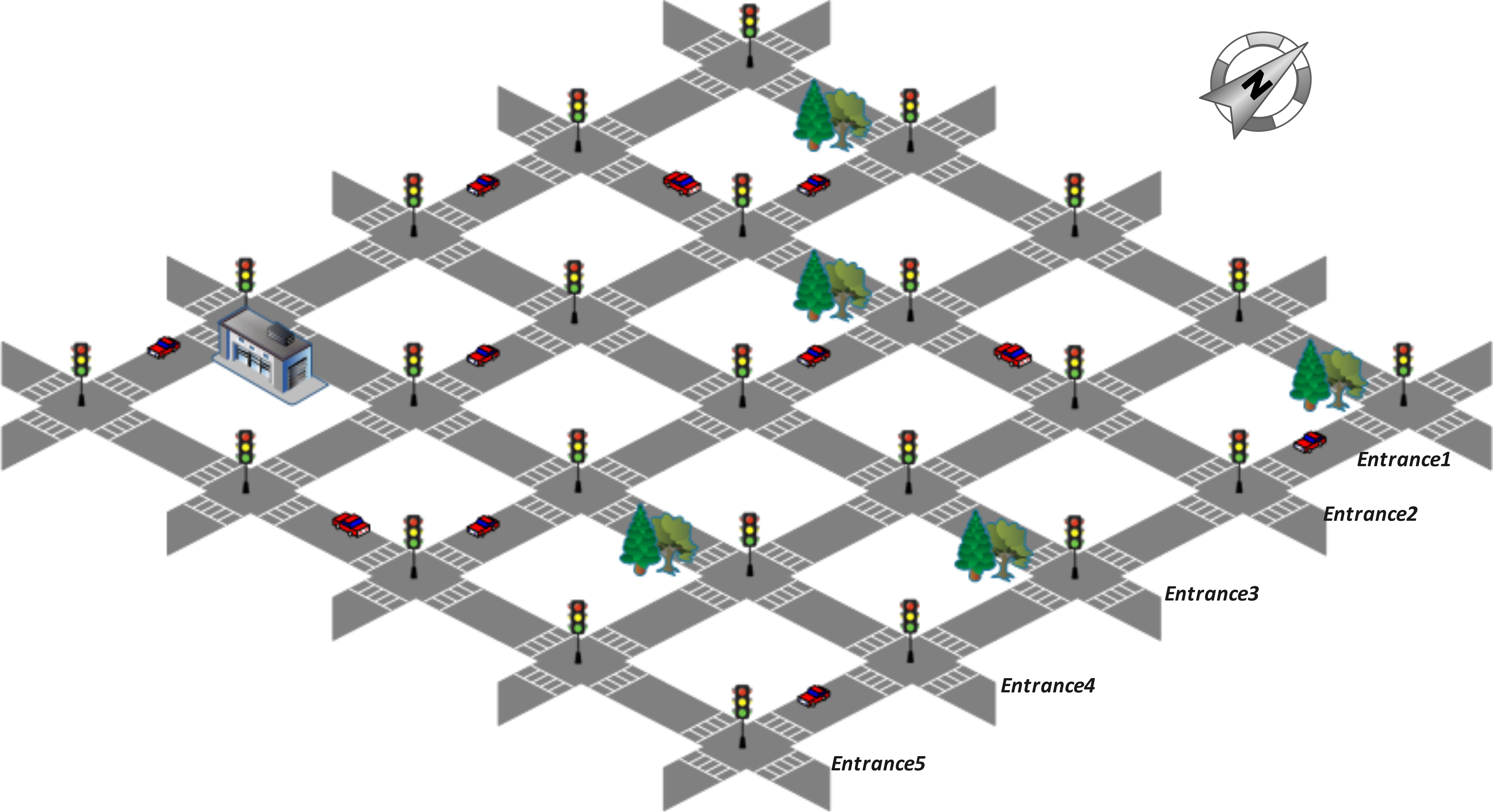}
	\caption{A road network with 25 traffic signal control agents.}
	\label{fig:4}
\end{figure}
The simulation road network is a $ 5 \times 5 $ as shown in Fig. \ref{2}. The total size of the grid is $ 600 \times 600 $ meters. The vehicle information of the road network, including input vehicle number (IVN)(vehicle number/episode) and traffic flow rate (TFR)(vehicle number/seconds) are summarized in Table \ref{ccccc}.   
\begin{table}[!]
	\centering
	\begin{tabular}{p{30 pt}p{40 pt}p{60 pt}p{50 pt}}
		\hline
		Direction & Entrance & IVN & TFR   \\
		\hline
		\multirow{2}*{North} & Entrance 1& 1000 veh/epi & 1/20 veh/s  \\

		& Entrance 2& 1000   & 1/10   \\
		& Entrance 3& 1100   & 1/15   \\
		& Entrance 4& 1050   & 1/20   \\
		& Entrance 5& 900   & 1/10   \\   
		\hline
		\multirow{2}*{South} & Entrance 1& 1100 & 1/15  \\
		
		& Entrance 2& 900   & 1/15   \\
		& Entrance 3& 900   & 1/20   \\
		& Entrance 4& 1000   & 1/10   \\
		& Entrance 5& 950   & 1/20   \\
		\hline
			\multirow{2}*{East} & Entrance 1& 1050 & 1/20  \\
		& Entrance 2& 1000   & 1/15   \\
		& Entrance 3& 950   & 1/15   \\
		& Entrance 4& 1000   & 1/20   \\
		& Entrance 5& 850   & 1/10   \\
		
		\hline
			\multirow{2}*{West} & Entrance 1& 950  & 1/15 \\
		
		& Entrance 2& 1000   & 1/20   \\
		& Entrance 3& 1050   & 1/10   \\
		& Entrance 4& 1000   & 1/10   \\
		& Entrance 5& 1000   & 1/20   \\
		\hline
	\end{tabular}
	\caption{\label{ccccc}The vehicle input information of the road network.}
	
\end{table}

\subsection{States}
\begin{figure}[!]
	\centering
	\includegraphics[width=0.6\linewidth]{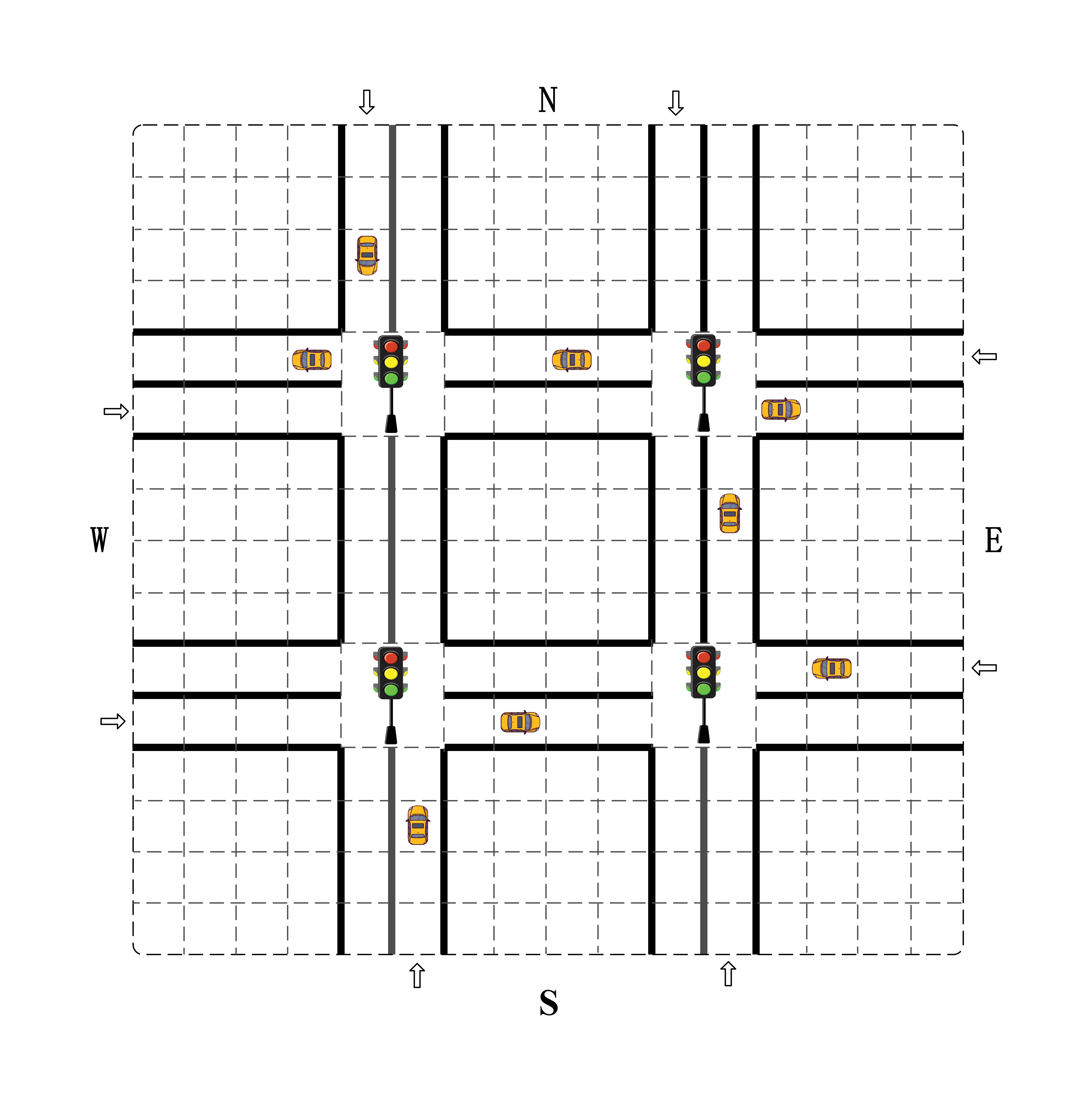}
	\caption{The snapshot of traffic at one time instant.}
	\label{chutian11}
\end{figure}
\begin{figure}[!]
	\centering
	\includegraphics[width=0.5\linewidth]{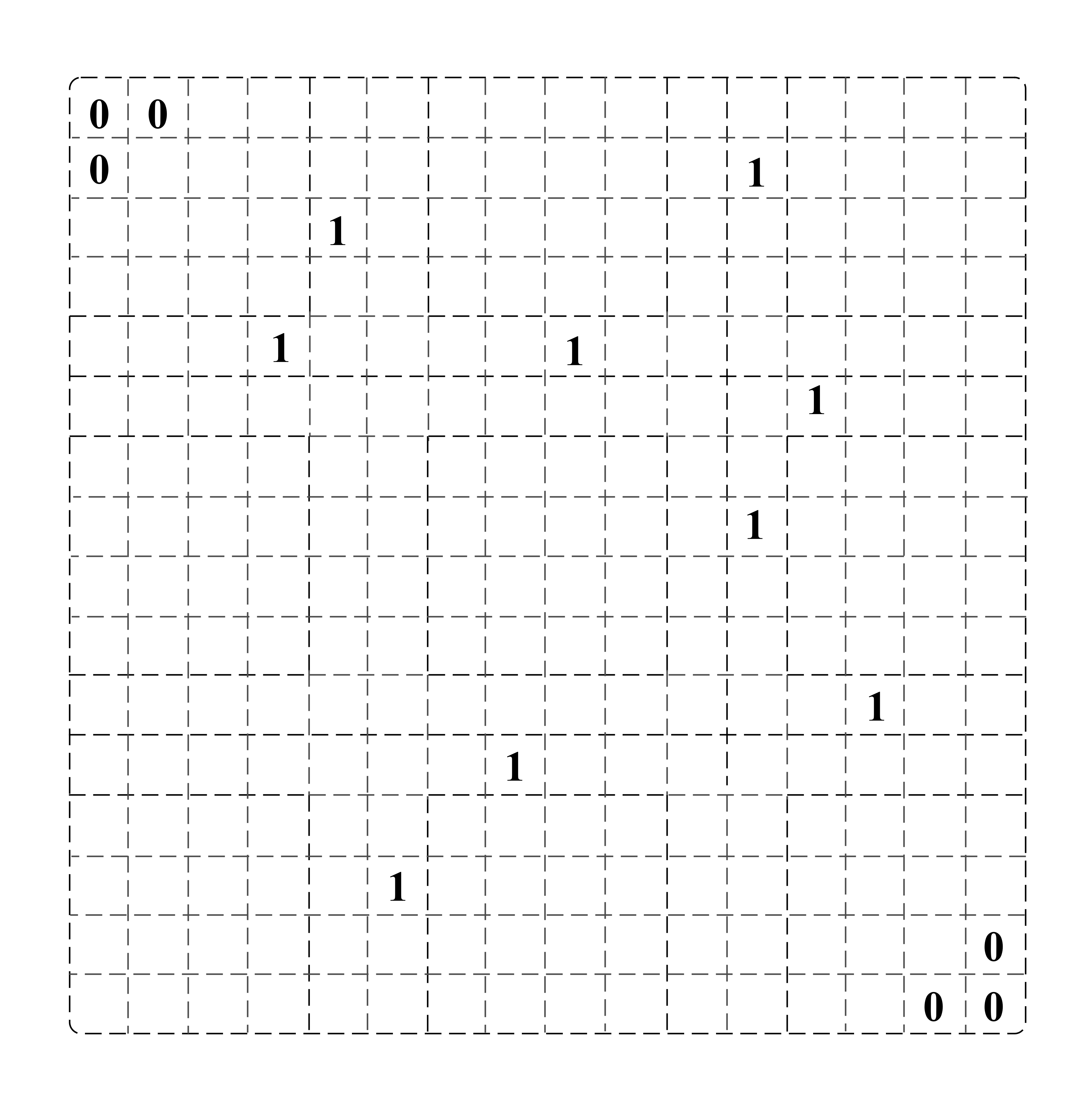}
	\caption{The corresponding position matrix of Fig. \ref{chutian11}.}
	\label{chutian22}
\end{figure}
Fig. \ref{chutian11} and Fig. \ref{chutian22} show how to set up the state values. Fig. \ref{chutian11} shows a snapshot of the traffic status at a simple network (four intersections for example), which is divided into square-shape grids. The position matrix has the same size of the grids, which is shown in Fig. \ref{chutian22}. The vehicle's position is set to 1, and the blank cells mean no vehicle in the
corresponding grid, which set to 0. It is worth noting that the length of each grid is equal to the length of the car, and a grid only have a 1 or 0.
\subsection{Actions}
In our model, the action space is defined by the green light phase duration of each stage change. We formulate a fixed period of 40 seconds. The green light duration of each traffic light can be changed to 10 seconds, 15 seconds, 20 seconds, 25 seconds, or 30 seconds. The remaining period is the duration of the red light (Each 40 seconds is an epoch, and after all vehicles of a route file run, the episode is incremented by 1 and the route file is reset). Each agent has the same action space.
\subsection{Reward}
The role of the reward is to provide feedback to the RL model about the performance of the previous action, so the formulation of the reward is particularly important. We define reward as the change in cumulative waiting time between two adjacent epochs. We define the cumulative waiting time of all vehicles in the $ T $ epoch as $ W_{T} $, then the reward in the $ T $ epoch is defined as:
\begin{equation}\label{key}
r_{T}=W_{T-1}-W_{T}
\end{equation}
The meaning of this reward is an increment in the cumulative waiting time of all vehicles after the action. If the reward in the current epoch is larger than before, it means that the waiting time of vehicles is decreasing, which can achieve the purpose of optimization.

\section{EVALUATION}
In this section, we will show the simulation results as well as the OPNDQN performance analysis for this research. We firstly illustrate the parameters used by the OPNDQN algorithm.
\subsection{Evaluation Parameters}
\begin{table}[H] 
	\normalsize
	\centering
	\caption{\label{tab:l1}PARAMETERS IN OFF POLICY NASH DEEP Q-NETWORK } 
	\begin{tabular}{lc}
		
		\hline Parameter & Value \\
		\hline Replay memory size $M$ & 20000 \\
		Minibatch size $B$ & 64 \\
		Starting $\epsilon$ & 1 \\
		Ending $\epsilon$ & $0.01$ \\
		Pre-training steps $t p$ & 2000 \\
		Target network update iteration & $100$ \\
		Discount factor $\gamma$ & $0.99$ \\
		Learning rate $\epsilon_{r}$ & $0.0001$ \\
		\hline
	\end{tabular}
\end{table} 
The model is trained in iterations. One iteration is an episode. The entire development environment is written in Tensorflow \cite{Marcham1929}, and the parameters of the model are shown in Table \ref{tab:l1}.
\subsection{Baseline Methods}
We compare the performance of the OPNDQN with the following baseline methods:

\textit{(1)} Multi-agent Q-learning (MAQL) \cite{abdoos2011traffic}: We use the Q-learning method separately at each intersection, the Q-table is applied to find the optimal traffic light control policy. Each intersection is independent, and no information sharing. 

\textit{(2)} Multi-agent Advantage Actor-Critic (MA2C) \cite{Chu2020}: We use A2C method separately at each intersection. Each agent uses critic network evaluates the policy of each actor and guides them to optimize their policies.

\textit{(3)} Fully Decentralized DQN Approach \cite{Mnih2015a}: We use the traditional DQN method separately at each intersection, each agent is essentially influenced by the latest actions of its neighbours, while searching the optimal strategy to control an intersection by a DQN.

\textit{(4)} Cooperative Multi-agent Deep RL Approach (Co-MARL): Haddad et al., proposed a Co-MARL approach in 2022 \cite{HADDAD2022105019}. The Co-MARL method applies a DQN, while transferring state, action and reward received from their neighbour agents to its own loss function during the learning process. Unlike our OPNDQN approach, Co-MARL does not let the agent play a game but uses the information transmitted by the neighbor agent as the state of learning to achieve the purpose of agent cooperation.
\subsection{Performance Analysis}
\begin{figure}[!]
	\centering
	\includegraphics[width=1\linewidth]{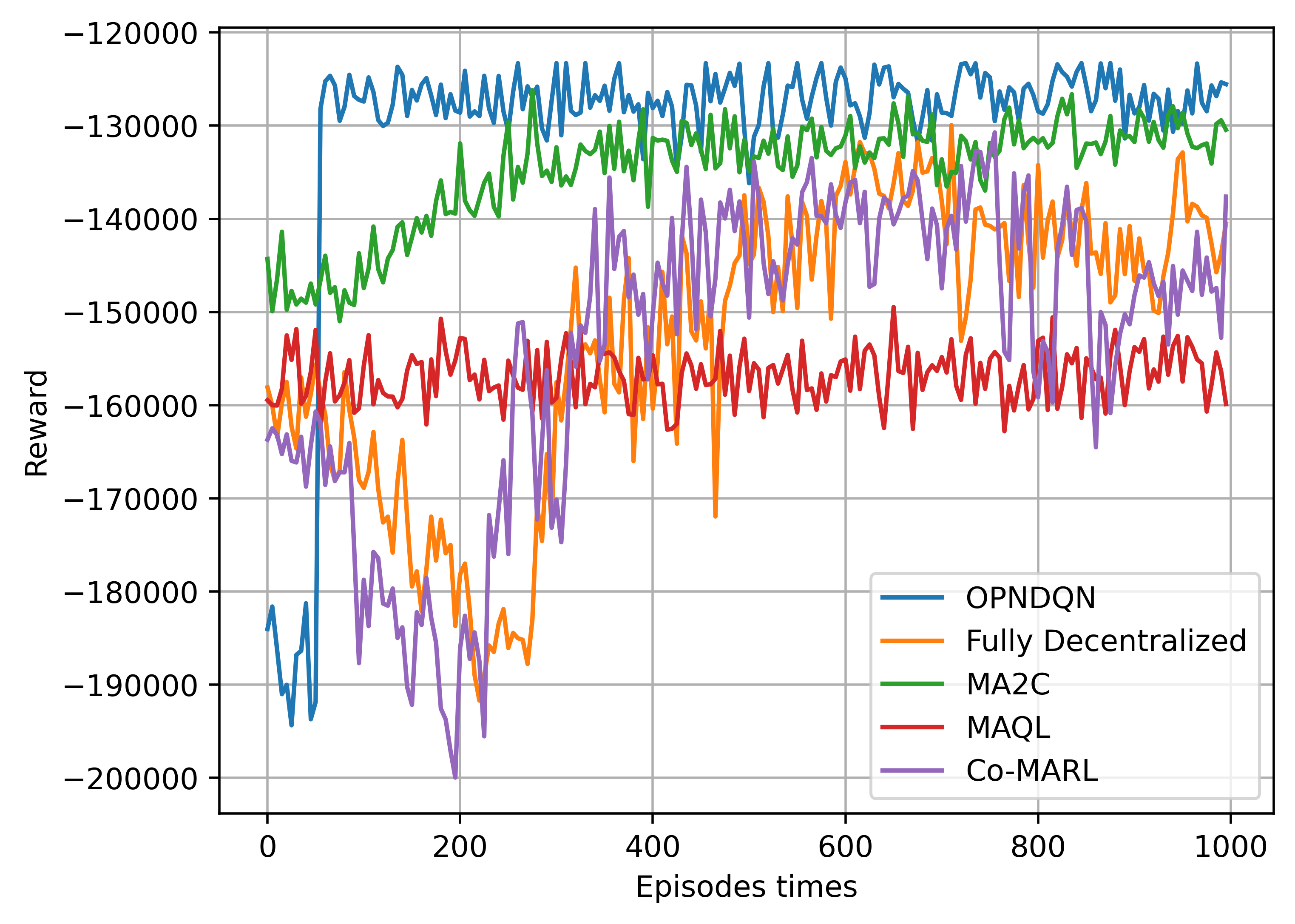}
	\caption{Performance comparisons for reward.}
	\label{temp1}
\end{figure}

\begin{figure}[!]
	\centering
	\includegraphics[width=0.9\linewidth]{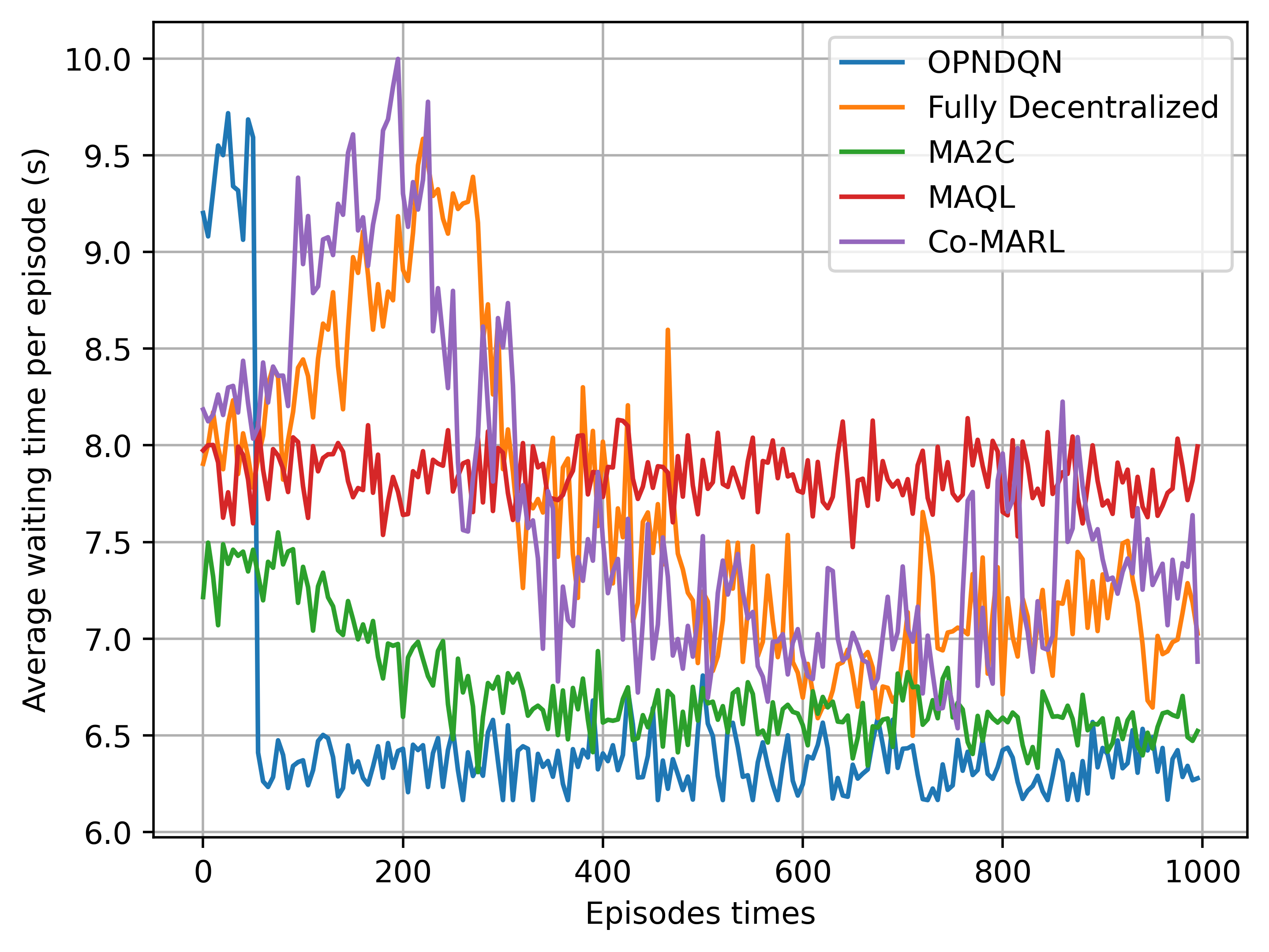}
	\caption{Performance comparisons for average waiting time.}
	\label{temp2}
\end{figure}

\begin{figure}[!]
	\centering
	\includegraphics[width=0.9\linewidth]{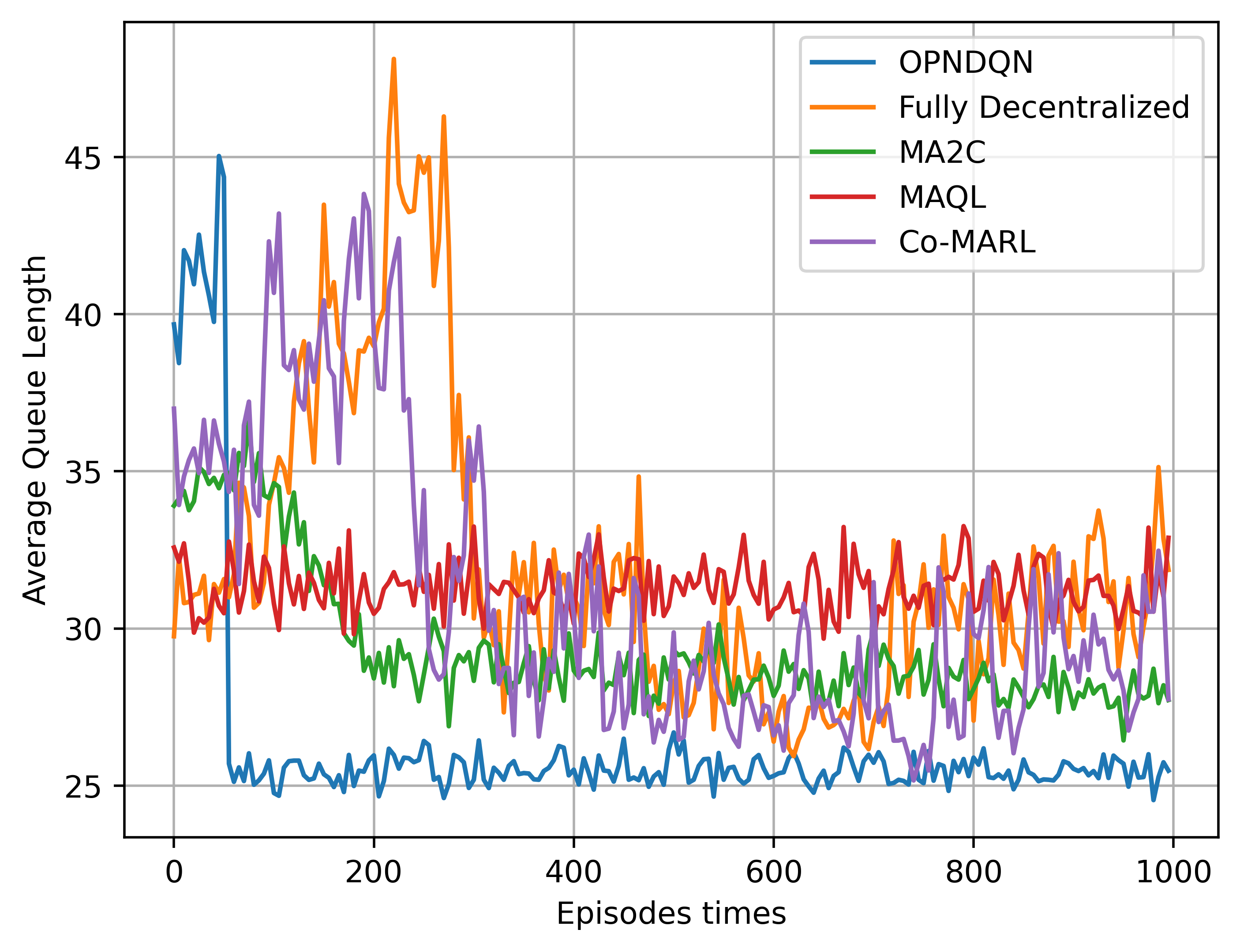}
	\caption{Performance comparisons for average queue length. }
	\label{temp3}
\end{figure}

The simulation results are shown in Fig. \ref{temp1}, \ref{temp2}, \ref{temp3}. Fig. \ref{temp1} shows the performance comparisons of cumulative reward in every episode under the same traffic flow rate. As we can see, when episodes approximately equal to 80, our method has converged. It takes 250 episodes for the fully decentralized DQN method, MA2C method and Co-MARL method to converge. The MAQL method is the least effective, and the model is almost impossible to converge. Meanwhile, in terms of training effect, the value of reward after convergence of our approach method is higher than the other four methods. Fig. \ref{temp2} shows five methods of average vehicle waiting time per episode. It can be seen that after the training of OPNDQN method, the average waiting time of vehicles is shortened to about 6.4 seconds, which is 0.2 seconds better than MA2C method, and 0.4 seconds lower than fully decentralized DQN method and MA2C method (MAQL was not taken into account, because it does not converge). Although in the average waiting time of vehicles, our method is not much more than the other methods, but in the convergence speed, our method is far better than the other four methods. To evaluate the performance of various signal control methods, we also compared the average queue length of vehicles at each intersection per episode in Fig. \ref{temp3}. We can see that OPNDQN is also performing very well, converging before 100 episodes and keeping the queue length at around 25. It can also be noted that the OPNDQN method has stronger stability after the completion of convergence. The three standard curves of fully decentralized method, MA2C method and Co-MARL method have large oscillation after convergence due to the mutual influence between intersections, which proves that our OPNDQN approach has better robustness and stability.


		
		

\section{CONCLUSION}
In this paper, we study the ATSC problem for a large traffic network, and then propose an OPNDQN approach. Firstly, we find a nash equilibrium between agents by using a fictitious game method. Secondly, we use the nash Q-value between agents rather than the maximum Q-value, it allows the agents to achieve equilibrium through learning. At the same time, OPNDQN method overcomes the problem that the fully centralized method is challenging to train large-scale neural networks. From the experimental results, the convergence speed of the OPNDQN algorithm is much higher than such approaches. Furthermore, our method has high stability after the algorithm converges. Experiments also demonstrate that with a large number of agents, the advantages of our method are prominent.

\end{document}